# Urea destabilizes RNA by forming stacking interactions and multiple hydrogen bonds with nucleic acid bases

U. Deva Priyakumar[†], Changbong Hyeon[‡], D. Thirumalai[*§], and Alexander D. MacKerell Jr.[*†]

[†]*Department of Pharmaceutical Sciences, School of Pharmacy, University of Maryland, Baltimore, Maryland 21201*

[‡] *Department of Chemistry, Chung-Ang University, Seoul 156-756, Republic of Korea*

[§] *Biophysics Program, Institute for Physical Science and Technology, University of Maryland, College Park, MD 20742*

E-mail: thirum@umd.edu;amackere@rx.umaryland.edu

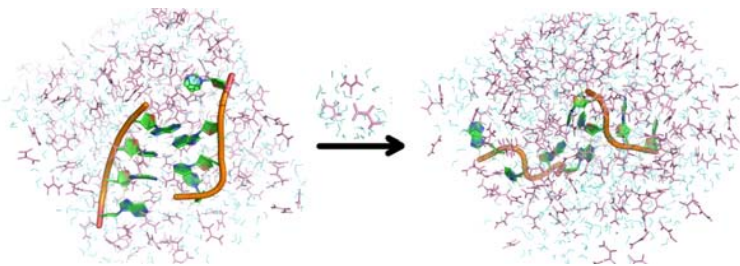

Urea titration of RNA by urea is an effective approach to investigate the forces stabilizing this biologically important molecule. We used all atom molecular dynamics simulations using two urea force fields and two RNA constructs to elucidate in atomic detail the destabilization mechanism of folded RNA in aqueous urea solutions. Urea denatures RNA by forming multiple hydrogen bonds with the RNA bases and has little influence on the phosphodiester backbone. Most significantly we discovered that urea engages in stacking interactions with the bases. We also estimate, for the first time, m-value for RNA, which is a measure of the strength of urea-RNA interactions. Our work provides a conceptual understanding of the mechanism by which urea enhances RNA folding rates.

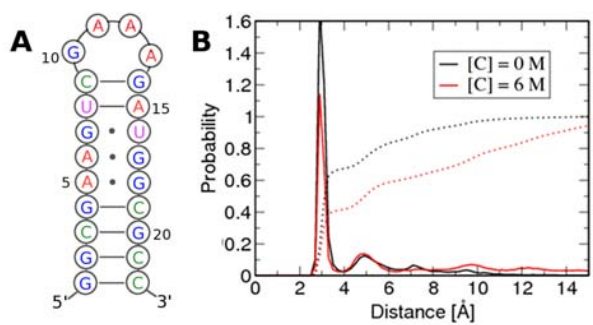

**Figure 1. Effect of urea on the P5GA hairpin. A.** Secondary structure map of P5GA. **B.** Probability distribution of the N1-N3, N1-N1 and N1-O2 inter-atomic distances of the GC(AU), GA and GU base pairs, respectively, in the stem. The dotted line is the integrated probability over the distances.

Urea has long been used to probe the stability and folding kinetics of proteins.[1] In contrast only recently it was shown that the RNA molecules that have a high propensity to misfold can be resolved using moderate amounts of urea.[2] Urea titrations can also be used to probe the interactions that stabilize the folded states of RNA.[2c] Although the mechanism by which urea denatures proteins is now fairly well understood[3] the nature of interactions by which urea destabilizes RNA is not known. In order to provide a microscopic basis for the action of urea on RNA we have carried out extensive all atom molecular dynamics (MD) simulations on two RNA constructs using two urea force fields. Destabilization of RNA is due to disruption of base-pair interactions by direct hydrogen bonding of urea with the bases. The simulations also reveal a novel mechanism in which urea molecules engage in stacking interactions with the purine bases.[4]

Analyses of 20 ns trajectories generated using MD simulations with a urea force field that was created as a part of the present work (see SI for simulation details, SI Figs. 1 and 2 and Tables 1-3 for urea parameter development, and for assessing the validity of the force field) of the 22-nucleotide RNA hairpin P5GA[5] (Fig. 1A) in various urea concentrations ([C]s) reveal that at high [C] the solvent-exposed stem regions lead to disruption of base pairing. The fraction of intact hydrogen bonds associated with the bases in the stem decreases from about 0.71 in the absence of urea to 0.46 in 8M urea. The loss of the Watson-Crick (WC) hydrogen bonds is accompanied by opening of the base pairs, which is reflected in the distribution of the hydrogen bond donor-acceptor distances ($R_{HB}$) in the hairpin stem (Fig. 1B). The base-paired state is indicated by a sharp peak at $R_{HB} = 3$Å, whose height decreases as [C] increases to 6M. The probability of sampling $R_{HB}$ distances that are greater than 10Å (Fig. 1B) increases greatly in high [C], which results in a rotation of the bases of the helix leading to N1-N3 distances of about 16Å.[6] Examination of opening at the individual base pair level reveals considerable heterogeneity[7] with the largest fluctuations occurring at the GA and GU mismatches. We also show that urea-induced disruption of the base opening due to the loss of WC hydrogen bonds is nonspecific in the sense that urea does not preferentially interact with a specific base pair. These finding suggests that denaturation of RNA is due to favorable non-specific interactions with amide-like surfaces of the nucleic acids. The average base-base interaction energies (GC, AU, AG, and GU) decrease substantially at high [C] (SI Table 4). When averaged over all base pair interactions in the stem the interactions become less favorable by about 2.7 kcal/mol at 6M relative to [C] = 0 (SI Table 4). The average interaction energies for certain base pairs (for example A6G17 and U8A15) are substantially less at high [C] relative to their values in water (see SI Table 4).

In contrast, the backbone conformational properties in the presence of urea are unperturbed, which is reflected in the distribution functions characterizing the phophodiester linkages. The angle distributions for $\alpha$, $\beta$, $\gamma$, $\chi$, and $\delta$ do not depend significantly on the urea concentration (Fig. 2 and SI Fig. 3). There are minor changes in the distributions of $\varepsilon$ and $\zeta$ change as [C] increases to 6M or 8M (Fig. 2). The small peak in the $\zeta$ distribution at 30° (Fig. 2) corresponds to the opening of the bases, which is in accord with previous studies that probed the base-flipping dynamics.[8] The [C]-dependent distributions of the $\zeta$ angle for each nucleotide show that the peak at 30° is also sampled by the GAAA tetra-loop (Fig. 1A). Taken together the results in Fig. 2 and SI Fig. 3 show that urea does not induce structural changes in the RNA backbone. The minor perturbations in the distributions of the $\chi$, $\varepsilon$, and $\zeta$ angles is merely a consequence of the opening of the bases.

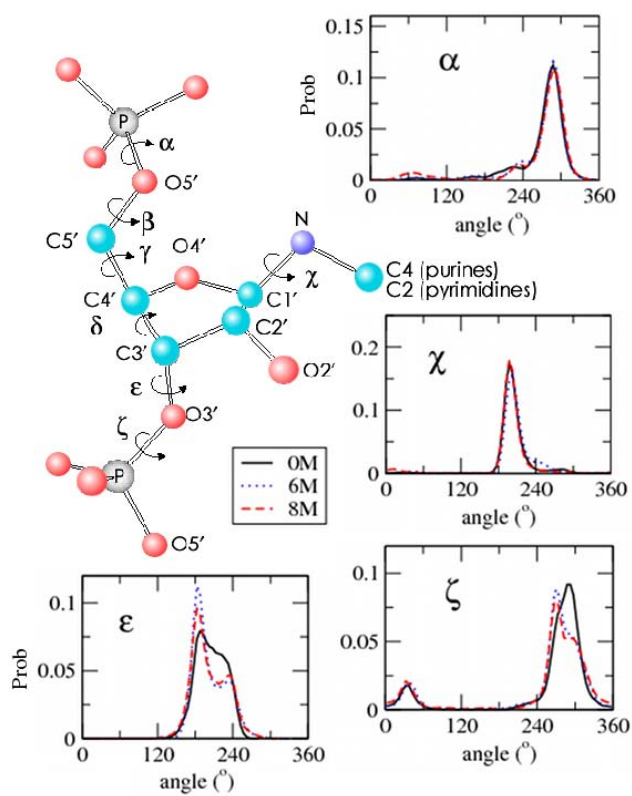

**Figure 2. Probability distributions of the dihedral angles along phosphodiester backbone of the RNA hairpin at [C]=0, 6, 8 M.**

To provide a molecular picture of urea-RNA interactions we introduce the dehydration ratio $\lambda_{DR} = \Delta N_W/N_U$ where $\Delta N_W$ is the difference in the number of water molecules in the first solvation shell of RNA as [C] increases from 0 and $N_U$ is number of urea molecules in the first solvation shell at [C] (SI Table 5). A value of $\lambda_{DR} > 1$ implies that more than one water molecule is exchanged for each urea. The values of $\lambda_{DR}$ change from 2.54 at 1M urea to about 0.85 at [C] = 8M. The decrease in $\lambda_{DR}$ at higher [C] is because the number of water molecules ceases to decrease while the number of urea molecules in the first solvation shell increases. The value of $\lambda_{DR}$ around the phosphodiester backbone is approximately independent of [C] (SI Table 5), which further indicates that the primary disruption of RNA structure is due to interactions of urea with the bases.

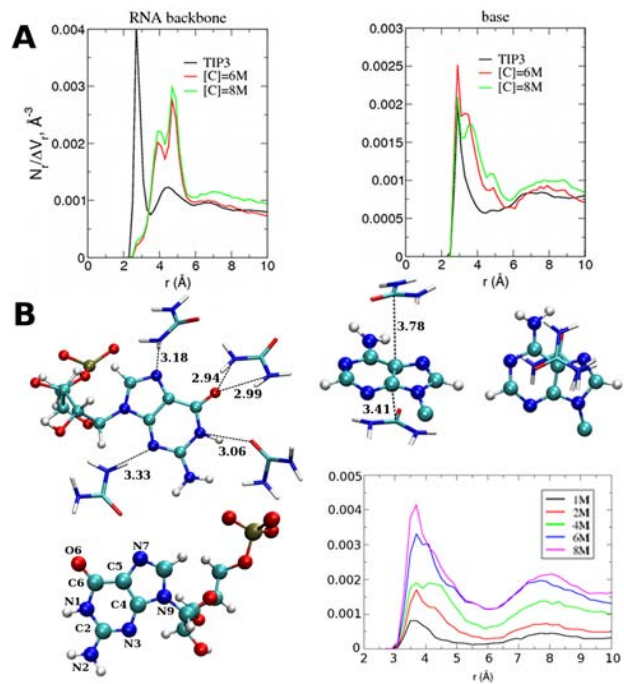

**Figure 3 A. RDFs of $O_U$ around the RNA nitrogen atoms at 6M and 8M urea. The water RDFs are scaled by 5 and 10 for the urea oxygen plots. Results are shown for the RNA backbone atoms (phosphodiester and sugar oxygen) and RNA bases. B. Structure with multiple hydrogen bonds between urea and RNA base and phosphate group. The panels on the right show the structure of urea-base stacking and the corresponding RDFs between the urea carbon and C4, C5, C2 (A and G) and C5, C6 (C and U) atoms. The contributions from individual atom are in SI Fig.13.**

The [C]-dependent values of $\lambda_{DR}$ also suggests that urea can engage in multiple interactions with the nucleic acid bases, which are reflected in the radial distribution functions (RDF) (Fig. 3A). There is an asymmetry in the interaction of $N_U$ and $O_U$ atoms of urea with RNA; $N_U$ atoms compete with water for direct hydrogen bonding interactions with both the bases and the backbone of the RNA as evident from a sharp peak in the RDF around 3 Å at all [C] (SI Fig. 4). The urea oxygen RDFs exhibit three distinct peaks approximately at 3, 4 and 5 Å (Fig. 3A). Surprisingly, the peak at 3 Å, which corresponds to the direct interactions with the RNA, is absent at [C] $\neq$ 0 indicating that there are only few direct interactions with the hydroxyl group of the ribose moiety and almost all the direct interactions occur with the bases. The additional peaks at approximately 4 and 5 Å correspond to oxygen atoms in urea molecules that indirectly interact with the RNA via the urea nitrogen atom. Representative examples of common hydrogen bonding interactions of urea with both the bases and backbone of the RNA show (Fig. 3B) that $N_U$ donates a hydrogen bond to N7 of a guanine base, with the distance between the hydrogen bond acceptor and the $O_U \approx 4$ Å corresponding to the second peak of the RDFs of the oxygen (Fig. 3A). Multiple urea-RNA interactions, leading to $\lambda_{DR} > 1$, include N7 and O6 of a single guanine base hydrogen bonding simultaneously the $N_U$ atoms (Fig. 3B), with the distances between the hydrogen bond acceptors and $O_U$ being around 5 Å, corresponding to the third peak in the RDFs (Fig. 3A). Remarkably, urea participates in stacking interactions with the bases (right panels in Fig. 3B), which further contributes to the destabilization of the folded RNA. Two urea molecules are positioned parallel to the purine base and the approximate interplanar distances are around 3.5 Å, which is comparable to the distance between the two rings in a benzene dimer that are stacked parallel to each other.[9]

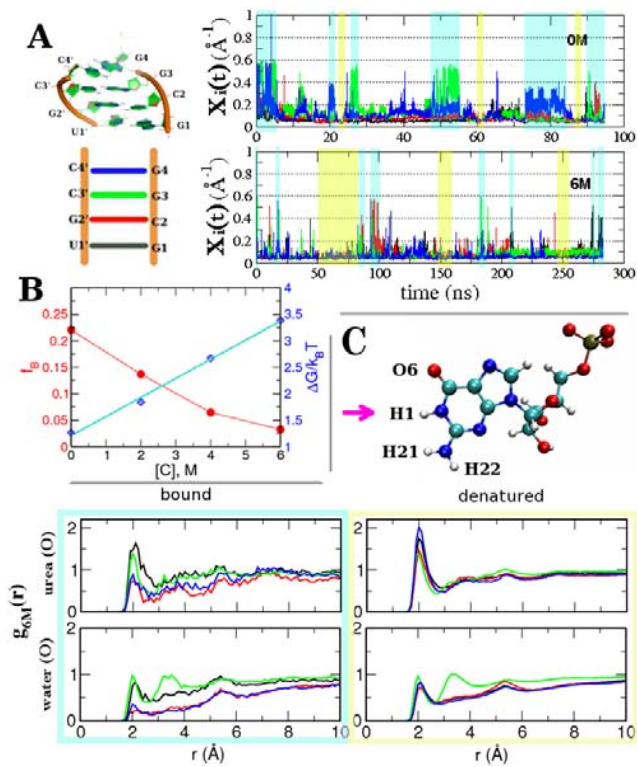

**Figure 4. Urea-induced structural transitions. A. The structure of RNA duplex (left). Inverse N1-N3 distances of the four base pairs (color-code, blue, green, red and black are used consistently) as a function of time at [C] = 0M (top) and [C[ = 6M (bottom); B. Fraction of bound $f_B$ and change in free energy ΔG as a function of urea concentrations. The fit is made for m-value analysis; C. RDFs of urea oxygen and water oxygen with respect to H1 atom of the G base when the duplex is bound or denatured at [C]=6 M.**

Because of the limitations in the sampling of the conformational space of the P5GA hairpin, we also simulated a smaller RNA duplex made of four complimentary base pairs at varying [C] (Fig. 4). To establish the robustness of the denaturation mechanism we used a different urea force field (see SI). Since the base pair distances ($r_i$) are subject to significant fluctuations when the base pairs are not formed, the inverse distance $X_i (=r_i^{-1})$ can be used to better visualize the equilibrium dynamics of the RNA duplex (Fig.4A). From 300 ns trajectories at each [C] (see Fig.4A and SI Fig.5) we calculated the fraction of bound duplex using $f_B = \tau_B/(\tau_B+\tau_U)$ and the change in free energy for (bound)⇔(unbound), $\Delta G[C]/k_BT=\log[(1-f_B)/f_B]=\log(\tau_U/\tau_B)$. The dwell time in the bound state $\tau_B$ satisfies $\Sigma_{i=1}^4 \Theta[X_c - X_i(t)] \neq 0$ and $\tau_U$ is the time for $\Sigma_{i=1}^4 \Theta[X_c - X_i(t)] = 0$; $\Theta(x)$ is the Heaviside step function, and $X_c = 0.2(Å)^{-1}$. For GC pair, $X_i$ is the inverse distance between H1 and N3 atoms in G and C, while for GU $X_i$ is taken between H1 in G and O2 in U (see SI Fig. 6). The decrease in $f_B$, relative to its value in water, as [C] increases (Fig. 4B), quantitatively demonstrates the destabilizing effect of urea on the RNA duplex. Because of the small size of the RNA the bound state is unstable even at [C]=0M, in accord with an estimated melting temperature in the range between 18-35 °C.[10]

Just as in proteins the free energy difference between the bound and the denatured states of the duplex RNA varies as $\Delta G[C]/k_BT = \Delta G[0] + m[C]$ (Fig. 4B). The value of m, which is the slope of the aqua line in Fig. 4B is ≈ 0.21kcal/mol·M. It is known that the m-value is a function of the RNA length, ion concentration, and depends sensitively on the valence of the counterion. Taking these factors into account, we find that the m-value obtained for the first time using simulations is in reasonable agreement with measurements on small duplexes.[2c,11]

The pair correlation functions involving water around the bound duplex were calculated using only those conformations that satisfy $\Sigma_{i=1}^4 \Theta[X_c - X_i(t)] =4$ (aqua shadow in Fig. 4A). The g(r) for the denatured state is calculated using the time traces in which the RNA duplex is fully denatured, i.e., $\Sigma_i \Theta[X_c - X_i(t)] =0$ (yellow shadow in Fig. 4A). The RDFs of $O_U$ and water oxygen relative to various atoms of the nucleic acid at varying urea concentrations (SI Figs. (**7-13**)) lead to a number of interesting conclusions: (i) Near the base, the density of water is below the bulk density (g(r→∞)≈1). As a consequence of the hydrophobic nature of the base the water distribution around the hydrogen atoms at H1 (Fig. 4C and SI Fig. **12**) and H21 or H22 of amide group of G (SI Fig. **11**) is g(r) < 1 regardless of the state of the RNA duplex. (ii) When paired bases are disrupted, $O_U$ forms a hydrogen bond with H1 atom in G (Fig. 4C and SI Fig. 7), which is reflected in the increase of RDF peak of urea at r = 2 Å. (iii) The g(r)s of urea or water around H21 or H22 (SI Fig. **8** and **11**), O6 oxygen in Guanine base (SI Fig. **9**), and OP1 or OP2 in phosphate group (SI Fig. **10**) are similar between bound and denatured forms. Thus, the disruption of the central hydrogen bond involving H1 of G, which is replaced by hydrogen bonds involving $O_U$, is the key event for the RNA denaturation. (iv) Comparison of g(r) functions in SI Figs. **8** and **11** shows in a dramatic fashion the depletion of water around the bases. More importantly, the ability of $O_U$ to form multiple hydrogen bonds is vividly illustrated (see Fig. **8** in the SI). (v) Stacking interactions with urea are reflected in the various RDFs (see Figs. 3 and SI).

Both sets of simulations show that destabilization of RNA is due to disruption of base-pair interactions by direct multiple hydrogen bonding with the bases and formation stacking interactions with the bases. In contrast to proteins, a multitude of favorable interactions largely involving the solvent-exposed bases leads to urea-induced destabilization of the structured RNA. In particular, there is no analogue of the stacking interactions involving urea in proteins, though stacking interactions in GdnHCl have been observed.[13] Finally, the proposed mechanism readily explains the observations[2] that urea-induced destabilization of base pair interactions in misfolded RNA molecules can increase the folding rates, thus acting as "chemical" chaperones.

**Acknowledgement.** This work was supported in part by grants from the National Science Foundation (CHE 09-10433), the NIH (GM51501), and the National Research Foundation of Korea (NRF) (R01-2008-000-10920-0, KRF-C00142, KRF-C00180 and 2009-0093817).

**Supporting Information Available:** Simulation methods and figures. This material is available free of charge in the Internet at http://pubs.acs.org.

**Supporting Information**

**Systems:** We used two RNA constructs in our simulations. The first is a 22 nucleotide (nt) hairpin, P5GA (PDB ID: 1EOR) whose structure has been determined using NMR. The other is an oligonucleotide consisting of two complimentary strands (duplex RNA) extracted from the full RNA structure (PDB ID: 1JP0). The smaller duplex RNA with only 8nts was chosen because its conformational space can be exhaustively sampled. The results obtained using both the systems further establish that our conclusions are robust.

**Urea force fields:** We developed a new urea force field for use in the P5GA (hairpin) simulations (see below). For the RNA duplex the parameters for urea were taken from Ref. 1. The excellent agreement in the proposed destablization mechanism between the different force fields lends further credence to the overall conclusions reached in this work.

**Urea Force Field for CHARMM:** We developed a new urea force field for simulations of the P5GA hairpin in aqueous urea solutions. Parameters for urea were optimized following the standard CHARMM protocols used previously in the simulations nucleic acids as well as other biological molecules[2,3]. Calculations were performed with the program CHARMM[4] or with the quantum mechanical program Gaussian03[5]. Briefly, internal parameters were optimized to reproduce survey data of the Cambridge Structural Database[6] or QM data (Table 1) and vibrational spectra obtained at the MP2/6-31G* level and scaled by 0.89 (Table 2)[7]. Charges and Lennard-Jones parameters were optimized to reproduce interactions of urea with water (Figure 2) obtained at the HF/6-31G* level on the MP2/6-31G* optimized gas phase geometry with the QM interaction energies scaled by 1.16 (Table 3). In addition, the free energy of aqueous solvation of urea, as calculated using the method of Deng and Roux[8], was used as target data for the optimization of the non-bonded terms. Individual simulations for the FE calculations included 5 ps of MD equilibration followed by 50 ps of sampling in a box of 125 TIP3P waters and 5 ps of equilibration and 20 ps of sampling in a box of 250 TIP3P waters.

Overall, the optimized empirical model is in good agreement with the target data. The level of agreement of the geometries is not optimal (Table 1) due to direct transfer of parameters from the CHARMM22 protein force field.[9] However, it was possible to obtain excellent agreement for the vibrational spectra including for the low frequency torsion and wag of the $NH_2$ moieties, which can undergo large distortions during MD simulations. Minimum interaction energies and distances of urea with water are in good agreement with the target QM data (Table 3). The non-bond parameters were adjusted to reproduce the most favorable interactions with the C=O and NH2 moieties with the interaction distances approximately 0.2 Å less than the target values as required to obtain the proper condensed phase densities. The final free energies of solvation were -13.3 and -13.1 kcal/mol from the perturbations in the boxes of 125 and 250 waters, respectively, are in good agreement with the target experimental value of -13.8 kcal/mol.[10] In addition, the dipole moment of the final empirical model was 4.88, overestimating the MP2/6-31G* value of 4.30, an overestimation required for the non-polarizable additive force field used in this study.

The validity of the current urea force field is further established by excellent comparison between the calculated and measured heats of sublimation for base crystals. Moreover, it has been shown that the method of using experimental data and quantum mechanical calculation to obtain force field parameters is accurate even in describing the configurations of benzene dimer.[11] In addition, we also tested the robustness of the urea-induced denaturation mechanism of RNA by performing simulation for the RNA duplex using an entirely different force field.

**Simulation Details:** Molecular dynamics (MD) simulations for P5GA were performed using the CHARMM program[4] employing the CHARMM27 nucleic acid force field[2] and the CHARMM modified TIP3P water model.[5] The P5GA hairpin was overlaid with a pre-equilibrated truncated octahedron of water or aqueous urea of varying concentrations. The solvent box was extended at least 9 Å beyond the non-hydrogen atoms of the RNA from all the sides. Water/urea molecules were removed if one or more of the solvent molecule's non-hydrogen atoms lie within 1.8 Å of the non-hydrogen atoms of the RNA. We added 21 sodium atoms to the resultant systems at random positions at a minimum of 3 Å from the RNA non-hydrogen atoms to maintain electrical neutrality. In all the subsequent minimizations and MD simulations, periodic boundary conditions were employed using the CRYSTAL[12] module in CHARMM. Energy minimizations were performed using the adopted basis Newton-Raphson (ABNR) method for 500 steps with mass-weighted harmonic restraints of 5.0 kcal/mol/Å on the non-hydrogen atoms of the RNA. After the initial minimization, each of these systems was subjected with a 20 ps MD simulation in the NPT ensemble followed by a 100 ps MD simulation in the NVT ensemble keeping the harmonic restraints. The short NPT simulation was carried out to allow the solvent molecules to move near the oligonucleotide and fill voids created by deleting the solvent molecules that overlapped with the RNA with the subsequent NVT simulation performed to allow full relaxation of the solvent, including the ions, around the RNA. Electrostatic interactions were treated using the particle mesh Ewald method.[13,14] Lennard-Jones interactions were truncated at 12 Å with a force switch smoothing function from 10 to 12 Å and the non-bond atom lists were updated heuristically. Production simulations were carried out at 278 K in accordance with the experimental conditions for 20 ns in the NPT ensemble with the Leapfrog integrator without any restraints. All the simulations employed an integration time step of 2 fs and the SHAKE algorithm[15] to constrain all covalent bonds involving hydrogen atoms. The NPT ensemble was achieved using Hoover chains[16] for temperature control and the Langevin piston method[17] was used to maintain a pressure of 1 ATM with a piston mass of 600 amu and the piston collision frequency set to 0. During the production run, coordinates were saved every 2 ps for analysis.

For the duplex RNA we used the NAMD molecular dynamics package with the CHARMM27 nucleic acid force field. The ds-RNA, with each strand consisting of 4nts, was solvated in a (42Å)$^3$ cubic box. The excess charges on the phosphate groups were neutralized with 11 Na$^+$ and 3 Cl$^-$ ions. To simulate the ds-RNA in aqueous urea solution we replaced water with 77, 154, and 230 urea molecules that results in 2M, 4M, and 6M urea solution. As in the simulations involving the P5GA hairpin we used periodic boundary conditions, and electrostatic interactions were treated using the particle mesh Ewald method. As in the study of P5Ga the simulations employed an integration time step of 2 *fs* and SHAKE algorithm to constrain all covalent bonds all the covalent bonds involving hydrogen atoms. The NPT ensemble was achieved using the Langevin thermostat with a friction coefficient of 5 $ps^{-1}$ on non-hydrogen atoms. Energy minimization that removes the instability of the entire system were performed for ds-RNA, urea, ions, and water respectively. The entire system was heated from 0 to 300 K every 1.2 nsec, after which the trajectories were generated for 80 ns in aqueous solution. The length of the trajectories at 2 M urea is ~ 260ns while at 4 M and 6M they were ~ 280 ns. The coordinates of the entire system were save every ps for analyses of the data.

**SI Table 1**: Internal geometries of urea

| Term | CSD | QM | Charmm |
|---|---|---|---|
| C=O | 1.26±0.02 | 1.233(1.267) | 1.225 |
| N-C | 1.33±0.03 | 1.381(1.3631) | 1.329 |
| N-C=O | 121.0±1.9 | 122.6(121.2763) | 124.0 |
| N-C-N | 117.9±2.0 | 114.8(122.0259) | 112.0 |
| N-C-N-H1 | | -168.3(-166.1) | -176.9 |
| N-C-N-H2 | | -25.0(-13.7) | -9.4 |

QM data from MP2/aug-cc-pVDZ optimized structure. Values in parentheses are values from an optimization of urea in the presence of 6 water molecules with the final geometry shown in Figure 1. Bond lengths are in Å and valance and dihedral angles are in degrees.

**SI Table 2**: Vibrational spectra of urea from MP2/6-31G* and final empirical models. MP2/6-31G* frequencies scaled by 0.89.

| | Scaled MP2/6-31G* | | | | | CHARMM | | | | |
|---|---|---|---|---|---|---|---|---|---|---|
| Mode | Frequency | Assignment | % | Assignment | % | Frequency | Assignment | % | Assignment | % |
| 1 | 151.3 | wNH2 | 51 | tNH2 | 47 | 151.1 | wNH2 | 70 | tNH2 | 29 |
| 2 | 382.6 | wNH2 | 81 | tNH2 | 20 | 352.0 | wNH2 | 78 | wCCO | 19 |
| 3 | 384.7 | tNH2 | 52 | wNH2 | 46 | 432.5 | tNH2 | 105 | | |
| 4 | 451.5 | dCCO | 78 | | | 450.8 | dCCO | 87 | | |
| 5 | 505.4 | tNH2 | 83 | | | 477.2 | tNH2 | 68 | wNH2 | 30 |
| 6 | 532.2 | rCCO | 83 | | | 583.7 | rCCO | 80 | | |
| 7 | 715.4 | wCCO | 108 | | | 687.6 | wCCO | 88 | wNH2 | 20 |
| 8 | 906.1 | sN-C | 86 | | | 882.2 | sN-C | 71 | | |
| 9 | 961.1 | rNH2 | 73 | sN-C | 25 | 1069.1 | rNH2 | 79 | sN-C | 18 |
| 10 | 1103.1 | rNH2 | 74 | sC=O | 17 | 1149.8 | rNH2 | 76 | sC=O | 22 |
| 11 | 1335.8 | sN-C | 52 | rNH2 | 17 | 1422.1 | dNH2 | 46 | sN-C | 26 |
| | | rCCO | 16 | | | | | | | |
| 12 | 1529.6 | dNH2 | 85 | | | 1580.8 | dNH2 | 89 | | |
| 13 | 1537.3 | dNH2 | 91 | | | 1643.9 | dNH2 | 49 | sN-C | 43 |
| 14 | 1674.8 | sC=O | 70 | | | 1779.1 | sC=O | 56 | sN-C | 20 |
| 15 | 3393.1 | sNH | 100 | | | 3401.4 | sNH | 99 | | |
| 16 | 3399.5 | sNH | 100 | | | 3415.3 | sNH | 99 | | |
| 17 | 3525.3 | sNHas | 100 | | | 3532.3 | sNHas | 99 | | |
| 18 | 3527.8 | sNHas | 100 | | | 3540.9 | sNHas | 100 | | |

Frequencies in cm$^{-1}$. Assignments represent the contribution of internal degrees of freedom to the potential energy distribution presented in percent contribution to each normal mode where s stands for bond stretching, d for valance angle deformations, r for rocking, t for torsions and w for the wagging mode.

**SI Table 3**: Minimum interaction energies and geometries between water and urea from QM and empirical model. Interaction orientations shown in SI Figure 2 and only the hydrogen bond distances were optimized.

| Interaction energy | QM1 | QM2 | EMP | Diff vs. QM2 |
|---|---|---|---|---|
| 1)C=O..HOH,linear | -6.73 | -6.93 | -7.63 | -0.70 |
| 2)C=O..HOH,120deg. | -8.61 | -9.31 | -9.27 | 0.03 |
| 3)N-H..OHH,Oside | -4.36 | -4.45 | -4.18 | 0.26 |
| 4)N-H..OHH,nonOside | -7.05 | -7.09 | -7.03 | 0.07 |
| Average Difference (QM2) | | | | -0.08 |
| RMS_Difference (QM2) | | | | 0.37 |
| Distances | QM1 | QM2 | EMP | |
| 1)C=O..HOH,linear | 2.02 | 2.01 | 1.74 | -0.27 |
| 2)C=O..HOH,120deg. | 1.96 | 2.01 | 1.72 | -0.29 |
| 3)N-H..OHH,Oside | 2.10 | 2.08 | 1.92 | -0.16 |
| 4)N-H..OHH,nonOside | 2.10 | 2.09 | 1.92 | -0.17 |

Energies in kcal/mol and distances in Å. QM interaction energies scaled by 1.16.

QM1: non-planar structure, QM2: enforced planarity. QM interaction energies are scaled by 1.16

Orientation 2) The C=O..H angle is fixed at 120˚.

**SI Table 4**: Energies of interaction[a] between the base pairs in the stem region of the P5GA RNA hairpin (see Fig. 1A in the main text for the secondary structure map of P5GA).

| Base Pair | 0M | 6M | 8M |
|---|---|---|---|
| **G2C21** | -13.6 ± 2.5 | -7.7 ± 2.4 | -14.2 ± 2.2 |
| **C3G20** | -17.7 ± 1.4 | -16.0 ± 1.2 | -10.4 ± 0.5 |
| **G4C19** | -21.4 ± 0.1 | -16.9 ± 1.8 | -21.8 ± 0.1 |
| **A5G18** | -9.8 ± 0.9 | -1.0 ± 0.5 | -11.7 ± 0.1 |
| **A6G17** | -10.2 ± 0.5 | -0.2 ± 0.0 | -2.7 ± 0.3 |
| **G7U16** | -5.0 ± 0.9 | -0.9 ± 0.3 | -1.8 ± 0.5 |
| **U8A15** | -6.9 ± 1.3 | -9.5 ± 0.8 | -0.7 ± 0.3 |
| **C9G14** | -18.1 ± 1.4 | -21.0 ± 0.5 | -12.5 ± 2.4 |
| **Ave[b]** | -12.9 | -9.2 | -9.5 |

(a) All values are given in kcal/mol and errors are the standard errors. (b) The average of all the average base pair interaction energies in a given system.

**SI Table 5**: Decomposition of the dehydration ratio[a] ($\lambda_{DR}$) at various structural elements of the P5GA hairpin as a function of urea concentration.

|  | 1M[b] | 6M | 8M |
|---|---|---|---|
| RNA | 1.94 | 1.11 | 1.26 |
| RNA backbone | 1.18 | 1.28 | 1.27 |
| RNA bases | 2.54 | 0.63 | 0.85 |
| Stem | 1.52 | 1.07 | 1.22 |
| Stem backbone | 1.16 | 1.29 | 1.28 |
| Stem bases | 1.67 | 0.48 | 0.71 |
| Major groove | 1.35 | 0.22 | 0.27 |
| Minor groove | 1.32 | 0.50 | 0.83 |
| Pyrimidines | 1.37 | 1.21 | 0.52 |
| Purines | 1.31 | 0.15 | 0.76 |
| Loop | 3.44 | 1.06 | 1.28 |
| Loop backbone | 0.98 | 1.12 | 1.16 |
| Loop bases | 6.76 | 0.85 | 1.21 |

(a) $\lambda_{DR}$ is a quantitative measure of the decrease in the number of water molecules to the increase in the number of urea molecules in the first solvation shell of the RNA as a function of urea concentration.
(b) Urea concentration.

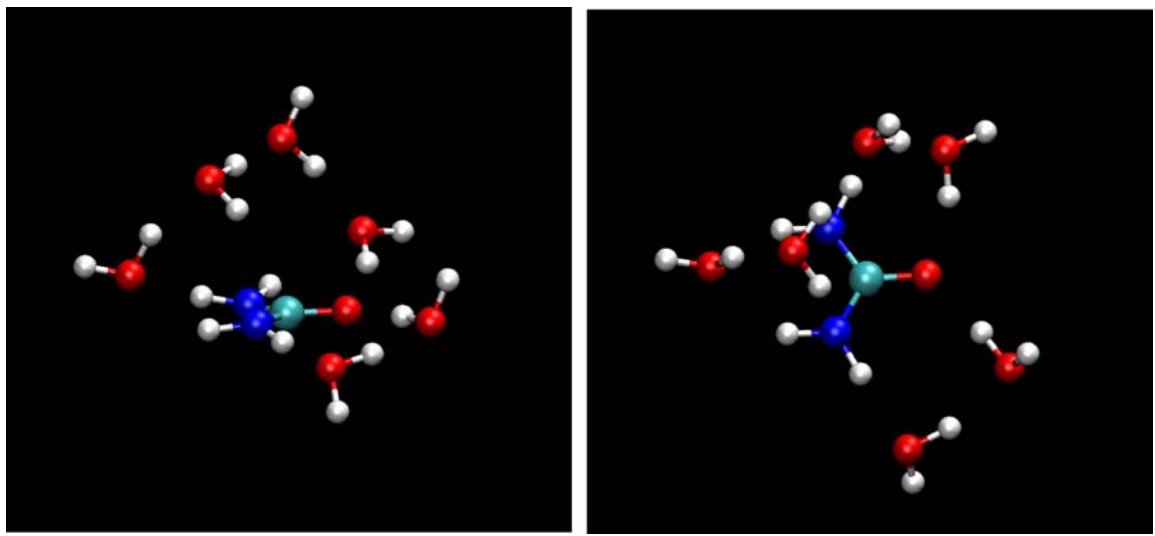

**SI Figure 1**: Images of a QM optimized structure of urea with 6 water molecules. The resulting configuration is used for analysis of the urea internal geometry.

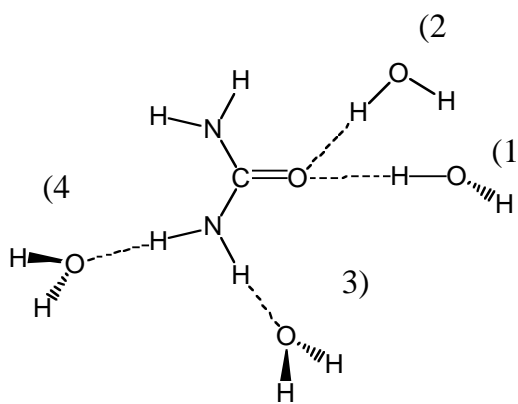

**SI Figure 2:** Diagram of interaction orientations between urea and water used in SI Table 3.

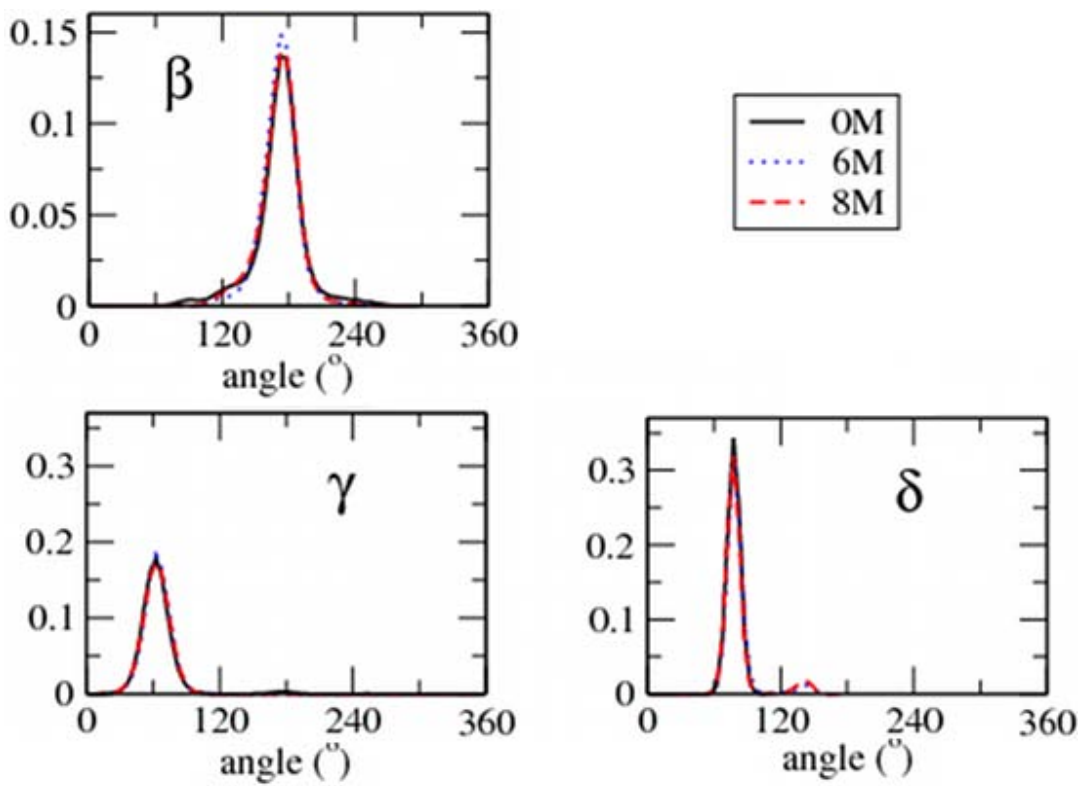

**SI Figure 3:** The probability distribution functions of selected phosphodiester-backbone dihedral angles for the P5GA hairpin (see Fig. 2 in the main text for definitions) at [C]=0, 6, 8M.

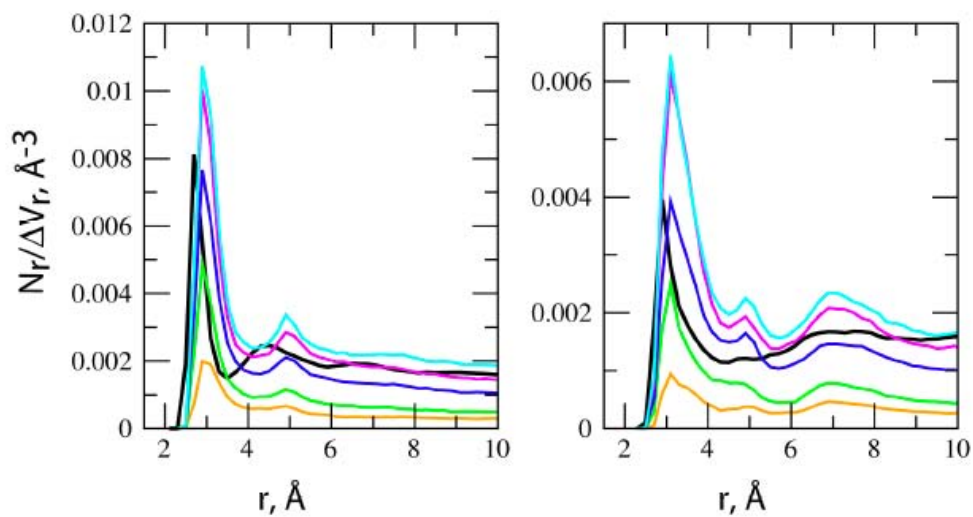

**SI Figure 4:** Radial distribution functions of the urea nitrogens oxygen. Results are shown for the RNA backbone atoms (left panel, includes phosphodiester and sugar oxygens) and the RNA bases (right panel). The color scheme is: 1M (orange), 2M (green), 4M (blue), 6M (magenta) and 8M (aqua) urea and for water at 0 M (black, bold). RDFs are normalized on a per atom basis. The water RDFs are scaled by 5 and 10.

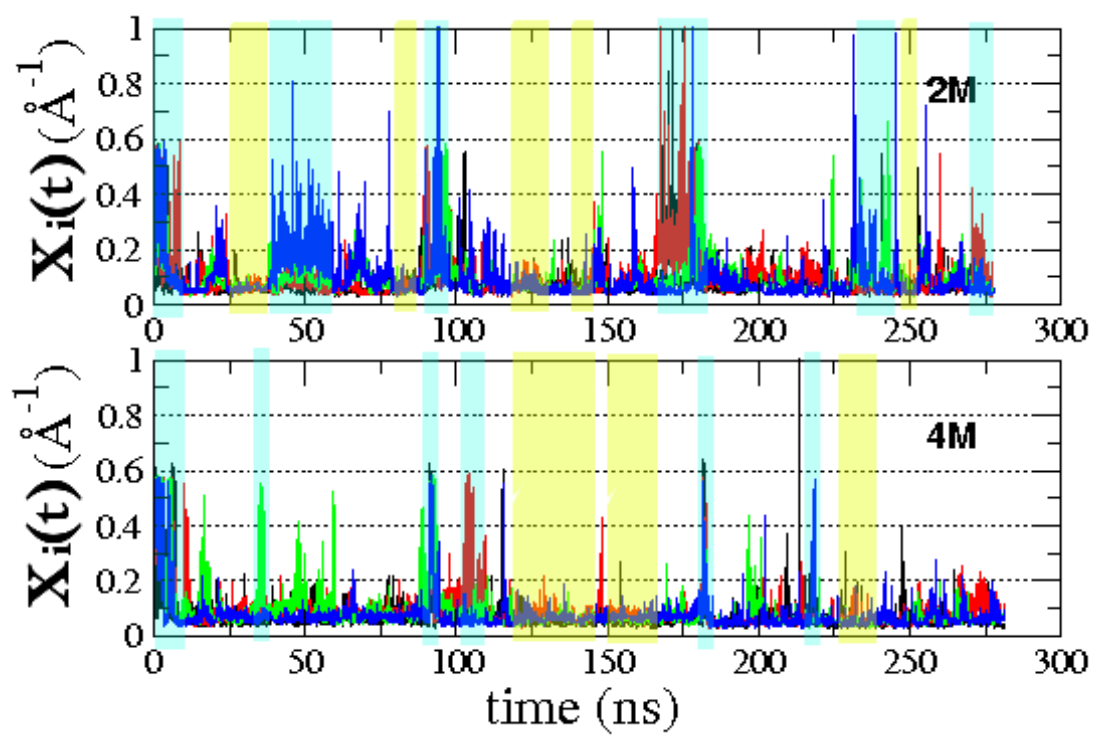

**SI Figure 5:** The base pair dynamics in the 8 nt RNA duplex measured using the inverse distance (1/r) are shown for [C]=2M and 4M urea.

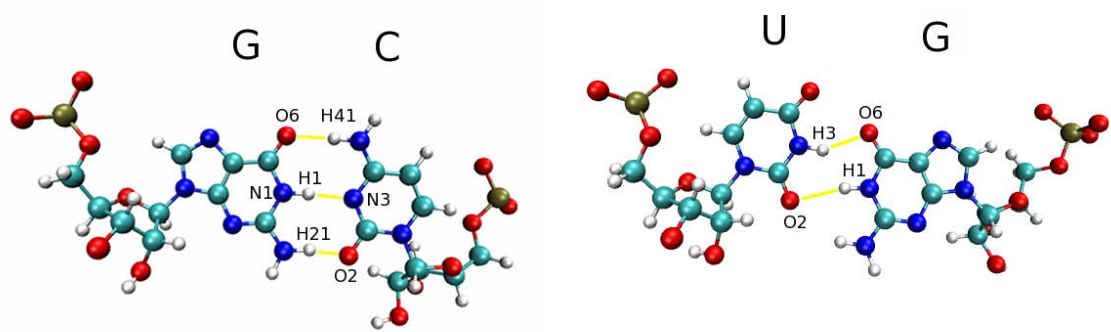

**SI Figure 6:** GC and UG base pairs and hydrogen bonds.

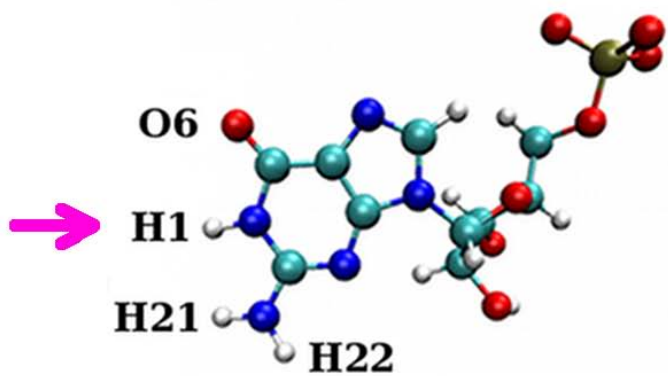
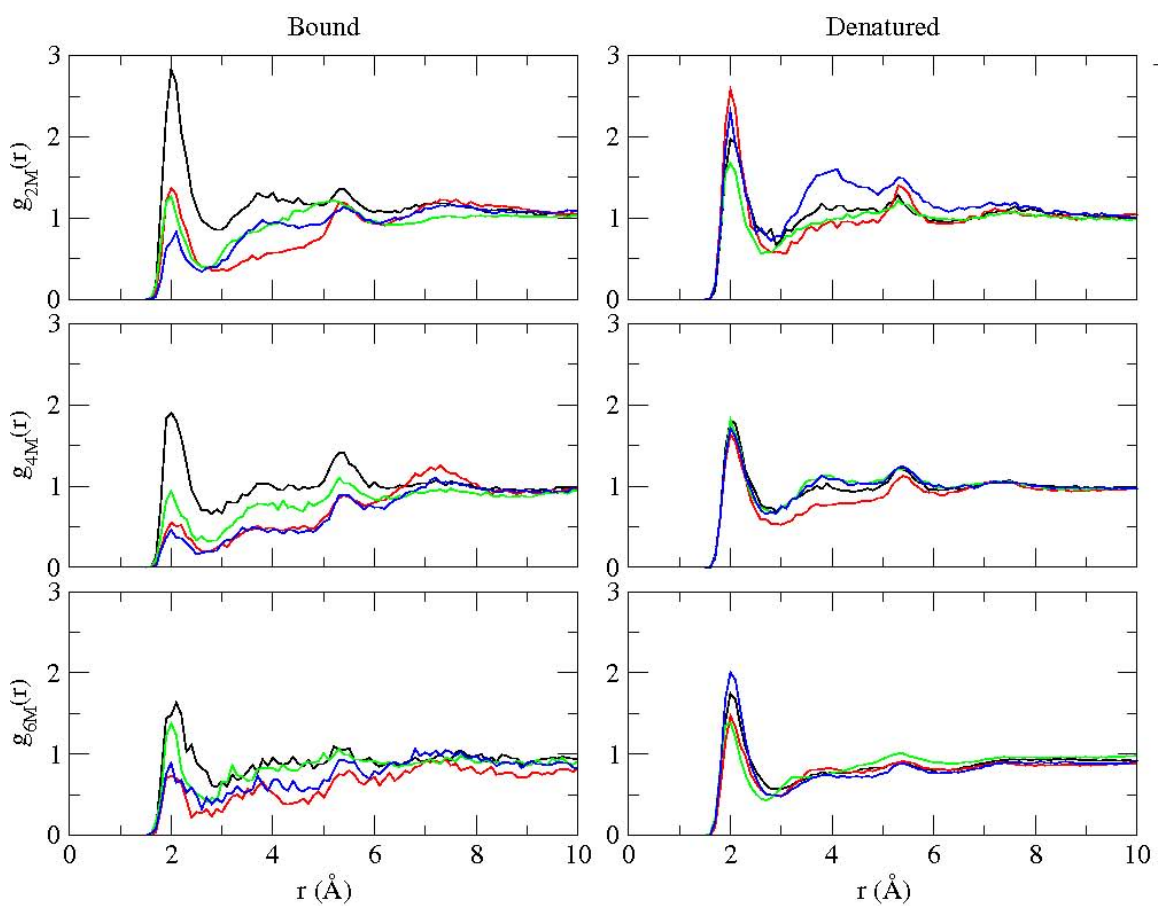

**SI Figure 7:** Pair functions for the urea oxygen with respect to H1 hydrogen of Guanine for the duplex RNA. Results in the panels on the left (right) are calculated using only the fraction of bound (unbound) RNA molecules.

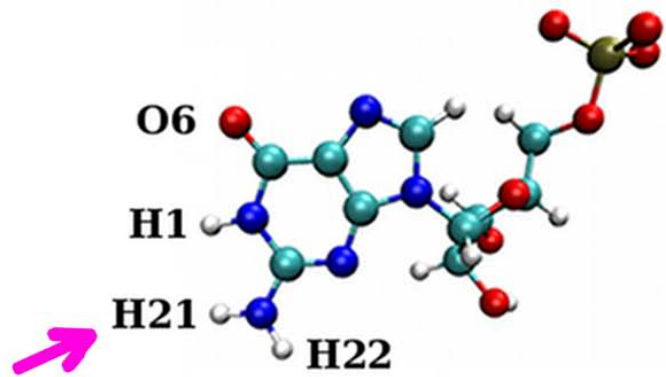

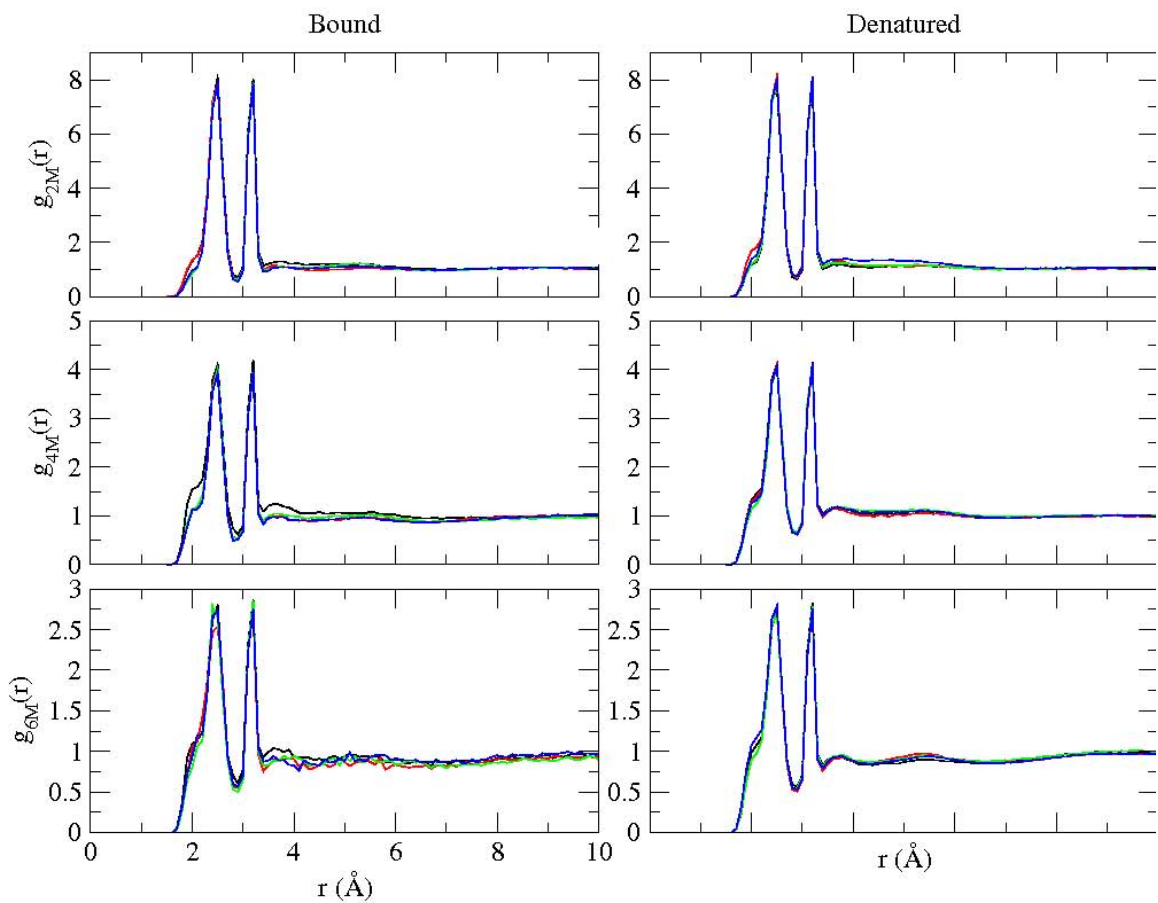

**SI Figure 8:** Pair functions for the urea oxygen with respect to H12 or H22 hydrogen of Guanine for the duplex RNA. Results in the panels on the left (right) are calculated using only the fraction of bound (unbound) RNA molecules.

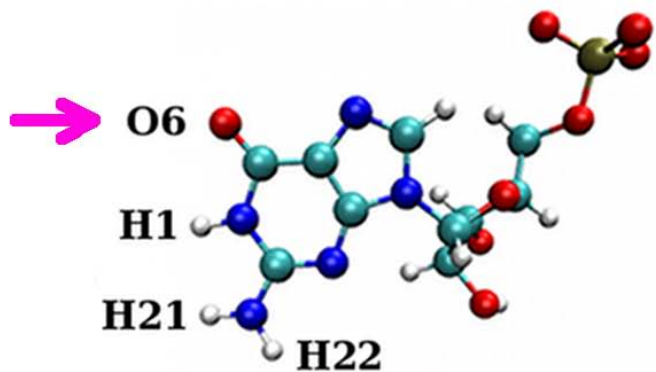

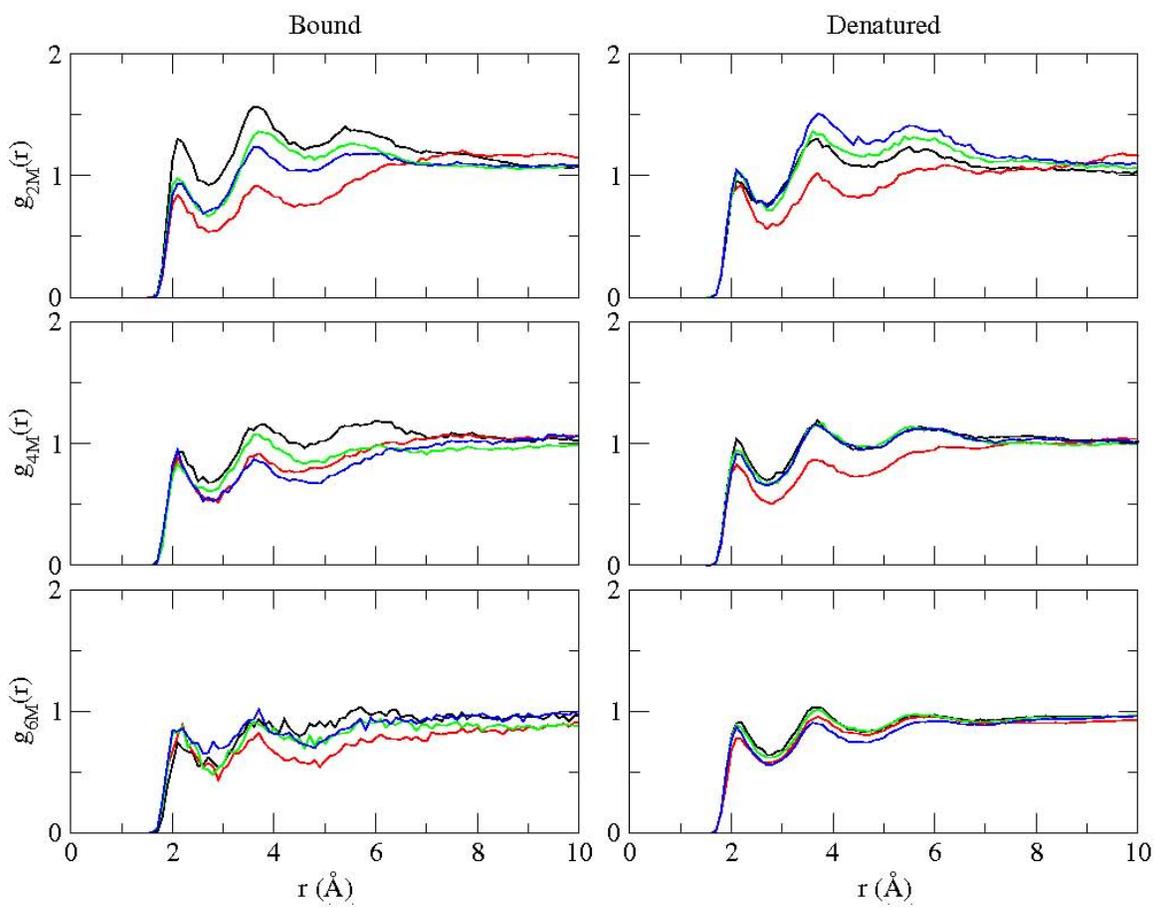

**SI Figure 9:** Radial distribution functions for amine hydrogen atom of urea with respect to O6 oxygen of Guanine at [C]= 2, 4, and 6 M in the duplex RNA. Results in the panels on the left (right) are calculated using only the fraction of bound (unbound) RNA molecules.

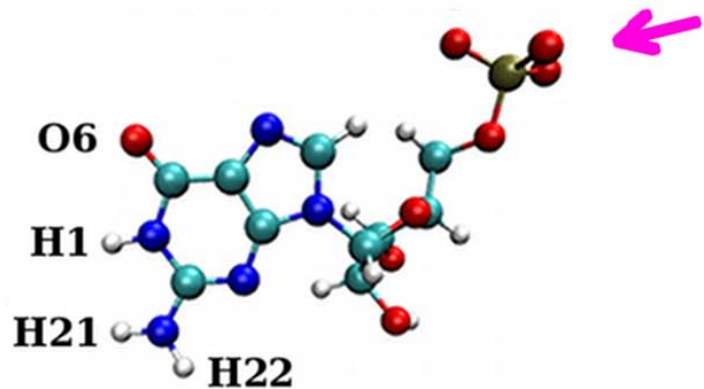
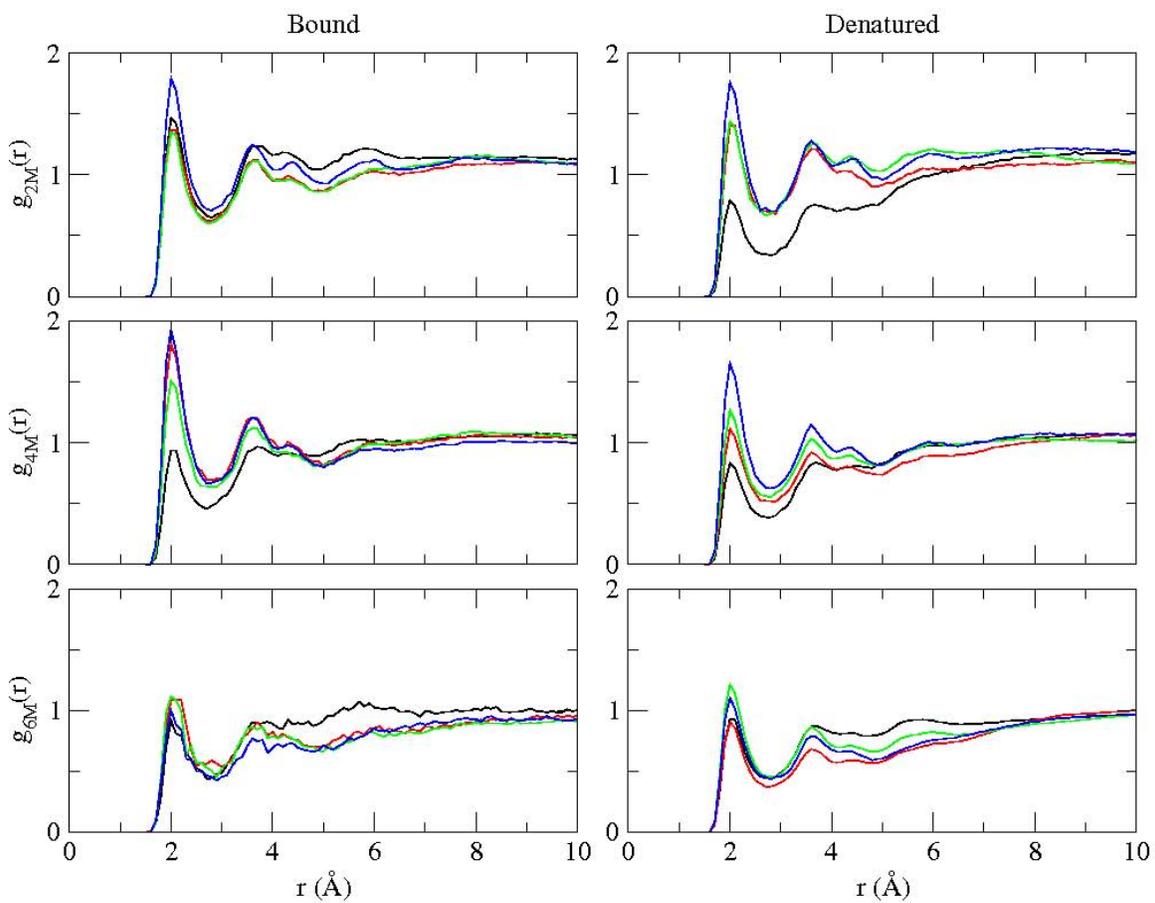

**SI Figure 10:** RDF for the hydrogens from amine group of urea with respect to the OP1 or OP2 oxygen of Guanine at [C] = 2, 4, and 6 M for the duplex RNA. Results in the panels on the left (right) are calculated using only the fraction of bound (unbound) RNA molecules.

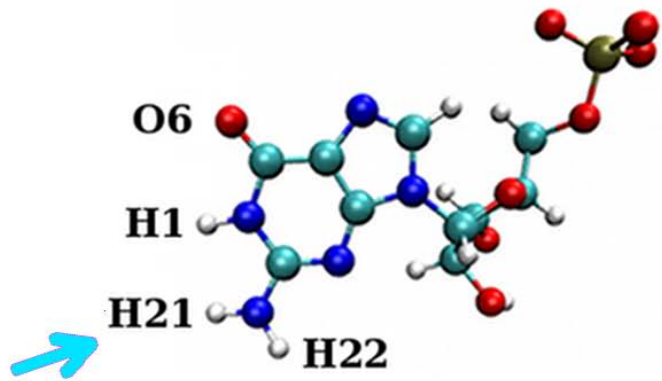

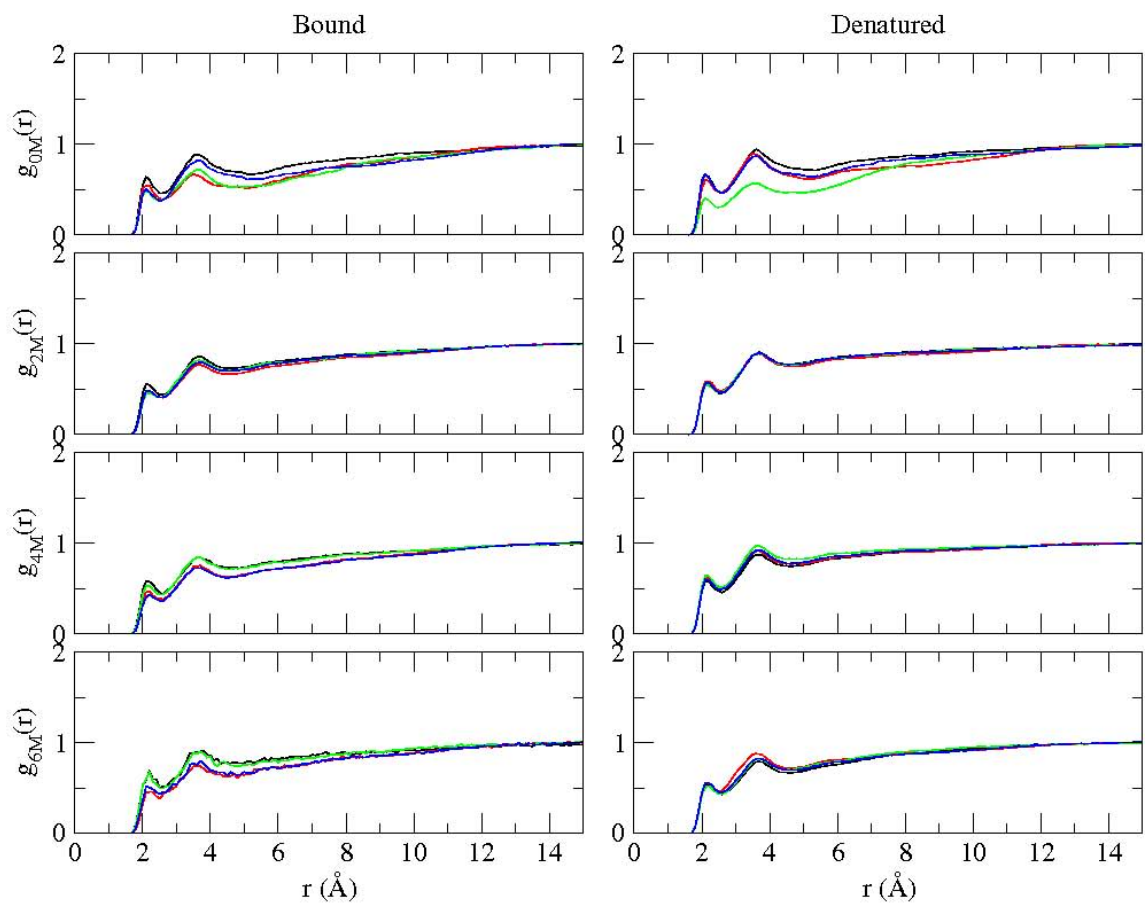

**SI Figure 11:** RDF for the water oxygen with respect to the H21 or H22 hydrogens of Guanine. Results in the panels on the left (right) are calculated using only the fraction of bound (unbound) RNA molecules.

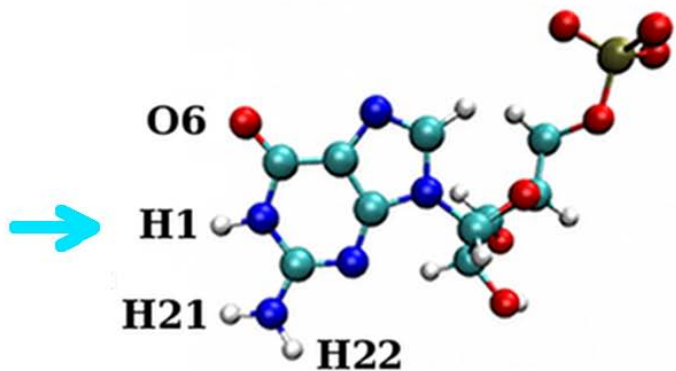

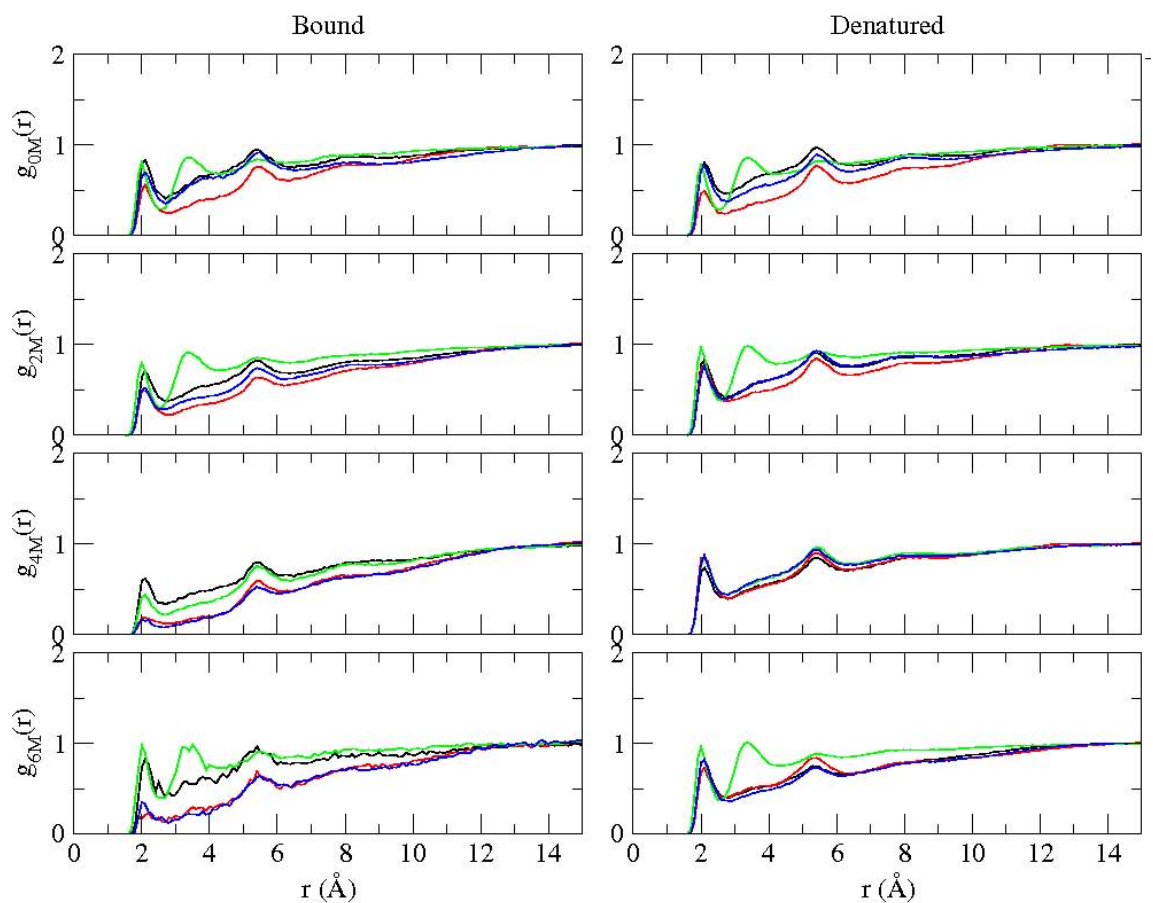

**SI Figure 12:** RDF for the water oxygen with respect to the H1 hydrogen of Guanine in duplex RNA at various concentrations of urea. Results in the panels on the left (right) are calculated using only the fraction of bound (unbound) RNA molecules.

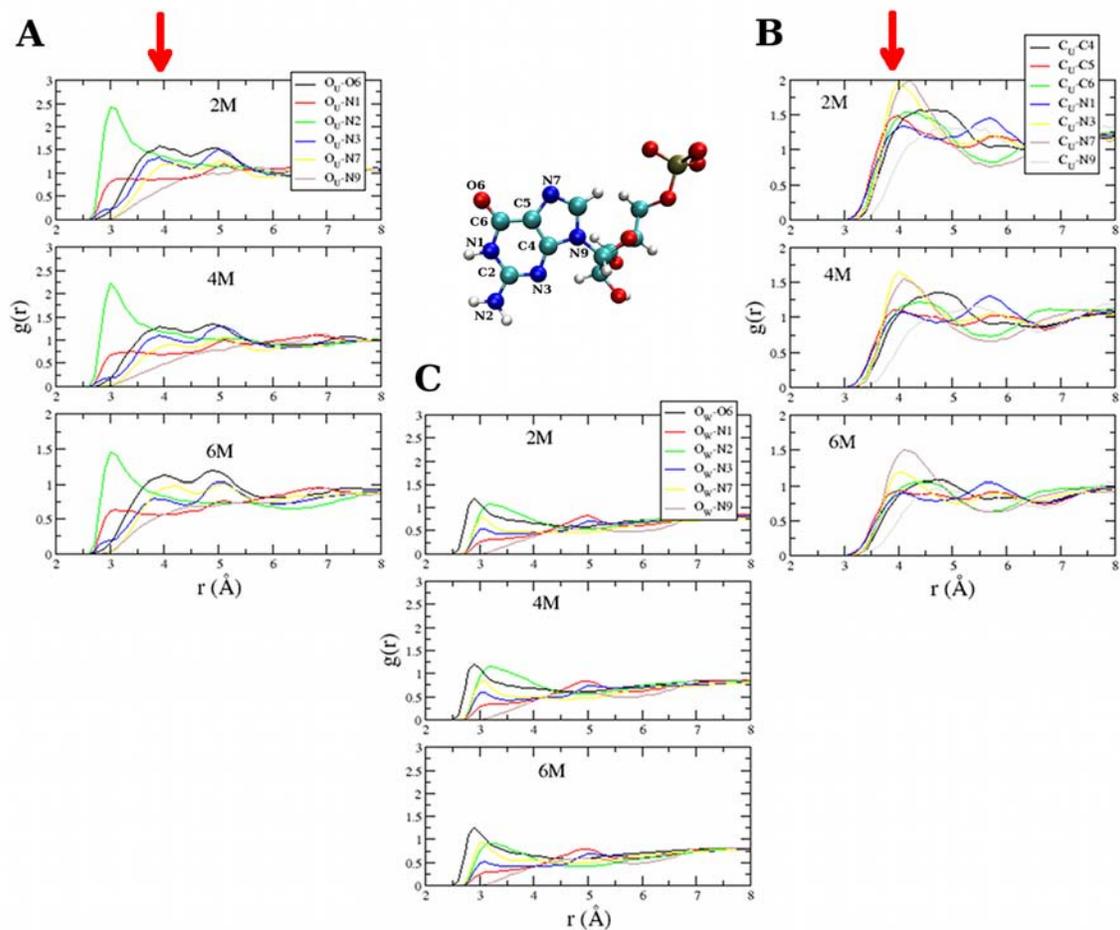

**SI Figure 13:** RDF for the urea oxygen ($O_U$) and urea carbon ($C_U$) with respect to various carbon, nitrogen and oxygen atom in Guanine base when RNA duplex is in denatured state for varying urea concentrations. The broad peaks at around 4 Å, indicated with red arrows in **A** and **B**, are due to the stacks formed between urea and base group. These results are consistent with the snapshot of the urea stack shown in Fig. 3B in the main text that were obtained using a completely different urea force field. For comparison, RDF for the water oxygen ($O_W$) with respect to the atoms in RNA base ring are shown in **C**, which confirms the depleted distribution of water due to the hydrophobic nature of the base.

22